\documentclass[letterpaper,twocolumn,showpacs,preprintnumbers,amsmath,amssymb]{revtex4-1}

\usepackage{graphicx}
\usepackage{dcolumn}
\usepackage{bm}

\DeclareGraphicsExtensions{.eps,.EPS}

\newcommand{\unit}[1]{\ensuremath{\, \mathrm{#1}}}

\begin{document}

\preprint{}

\title{
Energy shifts of Rydberg atoms due to patch fields near metal surfaces
} 

\author{J.~D.~Carter}
\author{J.~D.~D.~Martin}
\affiliation{%
Department of Physics and Astronomy and 
Institute for Quantum Computing \\
University of Waterloo, Waterloo, Ontario, N2L 3G1, Canada
}%

\date{\today}

\begin{abstract}
The statistical properties of patch electric fields due to a polycrystalline metal surface are calculated.  The fluctuations in the electric field scale like $1/z^2$, when $z \gg w$, where $z$ is the distance to the surface, and $w$ is the characteristic length scale of the surface patches. For typical thermally evaporated gold surfaces these field fluctuations are comparable to the image field of an elementary charge, and scale in the same way with distance to the surface.  Expressions for calculating the statistics of the inhomogeneous broadening of Rydberg-atom energies due to patch electric fields are presented.  Spatial variations in the patch fields over the Rydberg orbit are found to be insignificant.
\end{abstract}


\maketitle

\section{Introduction}

Excitation to a Rydberg state enhances an atom's interaction with a metal surface.  At large atom-surface distances, this results in energy level shifts that can be calculated using first-order perturbation theory \cite{PhysRevLett.68.3432}.  At smaller distances, the influence of the surface is more drastic---the Rydberg atom can be ``field-ionized'' by the surface \cite{chaplik:1968,PhysRevLett.85.5444}.  These phenomena may be visualized as arising from the interaction of the Rydberg atom with the electric fields due to its electrostatic ``image.''  Compared to an atom in the ground state, a Rydberg atom has an enhanced susceptibility to these fields.  This is because the Rydberg electron experiences a greatly reduced electric field from the ion core due to their larger average separation.

Polycrystalline metal surfaces generate inhomogeneous ``patch'' electric fields outside of their surfaces \cite{herring:1949}.  These fields may also influence Rydberg atoms, potentially causing both level shifts and ionization and competing with the more intrinsic image charge effects.  In general, patch fields arise from the individual grains of a polycrystalline surface exposing different faces of the bulk crystal.  Each face has a different work function due to differing surface dipole layers \cite{ashcroft:1976}.  For example, Singh-Miller and Marzari \cite{PhysRevB.80.235407} have recently calculated the work functions of the (111), (100), and (110) surfaces of gold and found 5.15, 5.10, and 5.04 eV, respectively.  These differing work functions correspond to potential differences just outside the surface beyond
the dipole layer.  Consequently, charge density must be redistributed on the 
surface to satisfy the electrostatic boundary conditions, producing
macroscopic electric fields \cite{ashcroft:1976}.  While patch fields were first discussed extensively in the context of thermionic emission \cite{herring:1949}, they may be present near polycrystalline metal structures of any type, including electrodes and electrostatic shields.

Recent advances in the trapping of cold atoms near surfaces have opened up the possibility of precision studies of Rydberg-atom--surface interactions as a function of atom-surface distance. For example, Tauschinsky et al.~\cite{PhysRevA.81.063411} have recently observed electromagnetically induced transparency due to Rydberg excitation of atoms at $10\unit{\mu m}$ to $200\unit{\mu m}$ away from a gold surface.   Lesanovsky {\it et al.}~\cite{PhysRevA.69.053405}  have calculated some interesting properties of Rydberg atoms exposed to inhomogeneous magnetic fields due to magnetic microtraps, and Crosse et al.~\cite{PhysRevA.82.010901} have recently calculated level shifts and transition rates of rubidium atoms near a copper surface at room temperature. In addition, there are several quantum information proposals that will involve Rydberg atoms in proximity to metal surfaces \cite{PhysRevLett.92.063601,PhysRevA.79.040304,PhysRevLett.93.103001}.
Consequently, it is desirable to be able to estimate the influence of 
patch fields on Rydberg atoms.  In this paper, we examine relevant models of the surface, report on the statistics of the patch fields, and determine the influence of these fields on Rydberg-atom energies.  We assume that the atom-surface distance is large compared to other relevant length scales and that the atomic energy level shifts can be treated using perturbation theory.

The model we adopt for the patch fields is similar to one used by Rzchowski and Henderson \cite{rzchowski:1988}.  Their work was motivated by the Witteborn-Fairbank experiment \cite{witteborn:1977}, which
was intended to compare the force of gravity on electrons and  
positrons.  Due to the relatively weak gravitational force, electrostatic shielding was necessary---the charged particles traveled down the
axis of a hollow copper cylinder used for shielding.  It was important to understand the variations in electrostatic potential along the axis of this tube due to patch fields, and Rzchowski and Henderson obtained results relevant to this geometry.  In the present work, we concentrate on a planar surface and the statistical properties of the electric field and its spatial derivatives.

\section{Rydberg-atom energy shifts in external fields}
We first calculate the energy-level shifts of a single atom in response to the local electrostatic potential $V(x,y,z)$ created by the patches. This will allow us to calculate the statistics of the energy shifts once the statistics of the patch fields are known.

We consider an addition to the atom's Hamiltonian 
$H_0$ of the form \cite{landauquantum:1981}
\begin{eqnarray}
\label{eq:h1}
H_{1} = && \sum_{i} \mu_{i} D_i V(x,y,z) \nonumber\\
&& + \frac{1}{6} \sum_{i,j} Q_{i,j} D_{i}D_{j} V(x,y,z) + ... \;,
\end{eqnarray}
where $\mu_i$ and $Q_{i,j}$ are the dipole and quadrupole moment operators, respectively, and $D_{i}$ is the operator representing the derivative with respect to the $i$th argument.  The quadrupole and higher-order moments will allow us to consider the influence of an electric field varying over the extent of the atom---of particular interest because Rydberg atoms are much larger than ground-state atoms.

We will consider the shift given in Eq.~(\ref{eq:h1}) using first-order perturbation theory.  In the absence of preexisting external fields, there would normally be no contribution from the dipole term to the first-order shift.  However, we will assume that a dc electric field aligned with the surface normal $\hat{z}$ has been applied. This may be done to enhance sensitivity to patch fields, break degeneracies, or for technical 
reasons (see, for example, Ref.~\cite{PhysRevLett.93.103001}).  The effect of this field is incorporated into  $H_{0}$.  Our basis states (eigenstates of $H_{0}$) will be considered to have a cylindrically symmetric charge distribution about the surface normal.  This symmetry constricts the moments, so that 
$\mu_x=\mu_y=0$, $Q_{xx}=Q_{yy}=-\frac{1}{2}Q_{zz}$, etc.

Equation (\ref{eq:h1}) involves the evaluation of arbitrary order mixed derivatives of the potential in all three spatial dimensions.  However, the introduction of cylindrical symmetry allows a considerable relaxation in this requirement if spherical (instead of cartesian) multipole moments are used.  Therefore, the external potential due to the patches is expanded in the form
\begin{equation}
\label{eq:serieseqlap}
V=\sum_{\ell,m} d_{\ell,m} r^{\ell} C_{\ell,m}(\theta,\phi)
\end{equation}
about the location of the atom, where $d_{\ell,m}$ are the expansion
coefficients, $r$ is the distance away from the center of the expansion, $\theta$ and $\phi$ the normal spherical coordinates with the polar axis aligned with the surface normal, and $C_{\ell,m}$ are rescaled spherical harmonics;   
$C_{\ell,m}=\sqrt{4\pi/(2 \ell +1)}Y_{\ell,m}$.

To obtain an expression analogous to Eq.~(\ref{eq:h1}), this new expansion of the potential is substituted into the volume integral for the electrostatic energy due to a charge distribution $\rho$ in an external potential, $E=\int d\tau \: \rho V$, with $d\tau$ as the differential volume element.  We obtain
\begin{equation}
E=\sum_{\ell} d_{\ell,0} \int d\tau \: \rho \: r^{\ell}P_{\ell}(\cos \theta ),
\end{equation}
where terms involving $m \neq 0$ do not appear, due to the cylindrical
symmetry of the charge distribution, and $P_{\ell}$ are
the Legendre polynomials ($=C_{\ell,0}$).  
The values of the $d_{\ell,0}$
coefficients in the expansion can be readily determined 
from Eq.~(\ref{eq:serieseqlap}) by evaluating
the derivatives of the potential with respect to the distance
to the surface $z$ evaluated at
the origin of the expansion, giving $d_{\ell,0}=(1/\ell!) D_z^{\ell}V(x,y,z)$, where $D_z^{\ell}$ means take the $\ell$th derivative with respect to the $z$ coordinate.  The first order energy shift can
be written in a form that only depends on the gradients of the field
in the $z$ direction evaluated at the location of the atom:
\begin{eqnarray}
\label{eq:firstorder}
E_{1} = \sum_{\ell} M_{\ell} [D_z^{\ell} V(x,y,z)],
\end{eqnarray}
where $M_{\ell}$ has been introduced to simplify notation.
To evaluate these matrix elements, we assume that only the charge distribution due to the Rydberg electron needs to be accounted for, so that
$M_{\ell}=(1/\ell !) q_e \langle \psi_0 | r^{\ell} C_{\ell,0} | \psi_0\rangle$, where $q_e$ is the electron charge, and 
$|\psi_0\rangle$ are the energy eigenstates of the zeroth order Hamiltonian
$H_0$.  The values of $M_{\ell}$ are proportional to the normal
spherical multipole moments (see, for example, Ref.~\cite{gray:1984}).

\section{Statistics of the Patch Fields}

As shown in the previous section, the energy of any particular atom depends on the field at its location. Consider an ensemble of atoms placed a certain distance $z$ away from the surface. In general, the patch fields are statistical in nature, so that spatial inhomogeneities in the field will cause an inhomogeneous broadening in the ensemble.  We can characterize this by the variance in the energy of a given state, calculated using Eq.~(\ref{eq:firstorder}), assuming that the average shift is zero:
\begin{eqnarray}
\label{eq:classical}
\langle (\Delta E)^2 \rangle_{C} = &&
\sum_{\ell,\ell^{\prime}} M_{\ell} M_{\ell^{\prime}} \nonumber\\ && 
\left\langle \Big[ D_z^{\ell} V(x,y,z) \Big] \left[ D_z^{\ell^{\prime}} V(x,y,z) \right] \right\rangle_{C},
\end{eqnarray}
where $\langle \cdots \rangle_C$ is used to specify an ensemble (classical) expectation value.

Therefore, to calculate the variance of atomic energy levels $\langle (\Delta E)^2 \rangle_C$ due to the statistical fluctuations in the field above the surface, we will develop expressions for
$\langle[D_z^{\ell} V(x,y,z)][D_z^{\ell^{\prime}} V(x,y,z)]\rangle_C$.
For example, the most important statistical fluctuation for Rydberg energy level shifts is of the electric field in the $z$ direction, which can be characterized by its root mean square (rms) value: $[\langle F_z^2\rangle-\langle F_z\rangle ^2]^{1/2}$, which is given by $\lbrace \langle [D_z^1 V(x,y,z)][D_z^1 V(x,y,z)] \rangle_C \rbrace^{1/2}$.

To calculate these statistical averages, we start by considering the solution of Laplace's equation $\nabla^2 V(x,y,z)=0$ above a plane surface when the potential on the surface is specified.  One particular solution of Laplace's equation is:
$V(x,y,z) = V_0 \: e^{ik_xx+ik_yy} \: e^{-kz}$, where $k=\sqrt{k_x^2+k_y^2}$ and $k_x$, $k_y$ and $V_0$ are constants.
Consider the following superposition of similar solutions (all integrations are assumed to run from negative to positive infinity, unless otherwise specified): \begin{equation} 
\label{eq:super}
V(x,y,z) = \int dk_x dk_y \: \tilde{V}(k_x,k_y) e^{ik_xx+ik_yy}e^{-kz} .
\end{equation} 
We may use this expression to determine the potential over any surface in the plane $z=0$ with a defined potential $V_s(x,y)$ by using the inverse Fourier transform to determine $\tilde{V}(k_x,k_y)$: 
\begin{equation} 
\label{eq:fourier}
\tilde{V}(k_x,k_y)  = \frac{1}{(2\pi)^2}\int dx dy \: V_s(x,y) e^{-ik_xx-ik_yy} .
\end{equation} 
Putting Eqs.~(\ref{eq:super}) and (\ref{eq:fourier}) together gives
\begin{eqnarray}
V(x,y,z) = && \frac{1}{(2\pi)^2} \int dk_x dk_y \:  e^{ik_xx+ik_yy}e^{-kz} 
\nonumber\\ && \times
\int  dx^\prime dy^\prime \: V_s(x^\prime,y^\prime) e^{-ik_xx^\prime-ik_yy^\prime} .
\end{eqnarray}

Consider the covariance between derivatives of the field evaluated at two points $a$ and $b$ in space, determined using the preceeding equation:
\begin{eqnarray}
\label{eq:firstmess}
&& \left\langle 
\Big[ D_p^{\ell} V(x_a,y_a,z_a) \Big] \Big[ D_q^{\ell^{\prime}} V(x_b,y_b,z_b) \Big]
\right\rangle_C  \nonumber\\ = &&
\frac{1}{(2\pi)^4} \int 
dk_{x,a} dk_{y,a}
dk_{x,b} dk_{y,b}
dx_a^\prime dy_a^\prime 
dx_b^\prime dy_b^\prime 
\nonumber\\ && 
\exp [
i k_{x,a}x_a 
+i k_{y,a}y_a
-k_a z_a
-i k_{x,a}x_a^\prime
-i k_{y,a}y_a^\prime \nonumber\\ &&
-i k_{x,b}x_b 
-i k_{y,b}y_b
-k_b z_b
+i k_{x,b}x_b^\prime
+i k_{y,b}y_b^\prime
] \nonumber\\ && \times
(\alpha_{p,a})^{\ell} (\alpha_{q,b}^*)^{\ell^{\prime}}
\times \langle
V_s(x_a^{\prime},y_a^{\prime}) 
V_s(x_b^{\prime},y_b^{\prime}) 
\rangle_C,
\end{eqnarray}
where 
$\alpha_{1,a}=i k_{x,a}$,
$\alpha_{2,a}=i k_{y,a}$, 
$\alpha_{3,a}=-k_{z,a}$,  
$\alpha_{1,b}=i k_{x,b}$,
$\alpha_{2,b}=i k_{y,b}$ and
$\alpha_{3,b}=-k_{z,b}$.  
As Eq.~(\ref{eq:classical}) shows, we only need $p=q=3$ (derivatives in the $z$ direction), but it is not difficult to deal with this slightly more general form, which will also allow us to calculate additional 
quantities, possibly of use to others, such as the total rms electric field.

We now make an assumption about the statistical nature of the field:  the correlation function $C$ for the surface potential only depends on the separation between the two points $a^\prime$ and $b^\prime$ (i.e., it is a ``stationary'' process):
\begin{equation}
\langle
V_s(x_a^{\prime},y_a^{\prime}) 
V_s(x_b^{\prime},y_b^{\prime}) 
\rangle_C \equiv
C(x_b^{\prime}-x_a^{\prime},y_b^{\prime}-y_a^{\prime}),
\end{equation}
and rewrite Eq.~(\ref{eq:firstmess}) using 
$\Delta x^\prime =x_b^{\prime}-x_a^{\prime}$,
and $\Delta y^\prime = y_b^{\prime}-y_a^{\prime}$:
\begin{eqnarray}
&& \left\langle 
\Big[ D_p^{\ell} V(x_a,y_a,z_a) \Big]
\Big[ D_q^{\ell^{\prime}} V(x_b,y_b,z_b) \Big]
\right\rangle_C = \nonumber\\ &&
\frac{1}{(2\pi)^4} \int 
dk_{x,a} dk_{y,a}
dk_{x,b} dk_{y,b}
dx_a^\prime dy_a^\prime 
d(\Delta x^\prime) d(\Delta y^\prime) 
\nonumber\\ && 
\exp [
i k_{x,a}x_a 
+i k_{y,a}y_a
-k_a z_a
-i k_{x,a}x_a^\prime
-i k_{y,a}y_a^\prime \nonumber\\ &&
-i k_{x,b}x_b 
-i k_{y,b}y_b
-k_b z_b \nonumber\\ &&
+i k_{x,b}(\Delta x^\prime+x_a^\prime)
+i k_{y,b}(\Delta y^\prime+y_a^\prime)
] \nonumber\\ && \times
(\alpha_{p,a})^{\ell} (\alpha_{q,b}^*)^{\ell^{\prime}}
\times C(\Delta x^\prime,\Delta y^\prime).
\end{eqnarray}
Use of the familiar relationship: \newline
$\int dx^\prime \exp [i x^\prime (k_a-k_b)] = 2\pi \delta ( k_a-k_b )$, where $\delta( \cdots ) $ is the Dirac $\delta$ function,  allows simplification to
\begin{eqnarray}
\label{eq:simpmess}
&& \left\langle 
\Big[ D_p^{\ell} V(x_a,y_a,z_a)\Big]\Big[D_q^{\ell^{\prime}} V(x_b,y_b,z_b)\Big]
\right\rangle_C 
= \nonumber\\ &&
\frac{1}{(2\pi)^2} \int 
dk_{x} dk_{y}
(\alpha_{p})^{\ell} (\alpha_{q}^*)^{\ell^{\prime}}
\nonumber\\ && \times
\exp [
i k_{x}(x_a-x_b) 
+i k_{y}(y_a-y_b) 
-k (z_a+z_b) ] 
\nonumber\\ && \times
\int
d(\Delta x^\prime) d(\Delta y^\prime) 
\exp [
i k_{x}(\Delta x^\prime) 
+i k_{y}(\Delta y^\prime)]  \nonumber\\ && \times
C(\Delta x^\prime,\Delta y^\prime).
\end{eqnarray}
Assuming the surface has no preferred direction,
$C(\Delta x^\prime,\Delta y^\prime)$ is only a function of 
$\Delta r^\prime=\sqrt{{\Delta x^\prime}^2+{\Delta y^\prime}^2}$,
and the evaluation of the last integral in Eq. \ref{eq:simpmess} is equivalent to taking the two-dimensional (2D) Fourier transform of a radially symmetric function (see, for example, Ref.~\cite{baddour:2009}): 
\begin{eqnarray}
\label{eq:wk}
W(k) \equiv && \frac{1}{2\pi} \int
d(\Delta x^\prime) d(\Delta y^\prime) 
\exp [
i k_{x}(\Delta x^\prime) 
+i k_{y}(\Delta y^\prime)]  \nonumber\\ && \times
C(\Delta x^\prime,\Delta y^\prime) \nonumber\\  
= && \int_{0}^{\infty} d(\Delta r^\prime) 
\Delta r^\prime J_0(k \Delta r^\prime)
C(\Delta r^\prime),
\end{eqnarray}
where $J_0(\cdots)$ is the zeroth-order Bessel function. Equation (\ref{eq:simpmess}) may then be written as
\begin{eqnarray}
\label{eq:mainresult}
&& \left\langle 
\Big[ D_p^{\ell} V(x,y,z) \Big]
\Big[ D_q^{\ell^{\prime}} V(x+\Delta x,y+\Delta y,z+\Delta z) \Big]
\right\rangle_C  \nonumber \\ && =
\frac{1}{2\pi} \int 
dk_{x} dk_{y}
(\alpha_{p})^{\ell} (\alpha_{q}^*)^{\ell^{\prime}} W(k)
\nonumber\\ && \times
\exp [
i k_{x}\Delta x 
+i k_{y}\Delta y 
-2 k z -k \Delta z ] .
\end{eqnarray}
A generalisation of this result to mixed derivatives is
straightforward, but the notation cumbersome.  To evaluate
Eq.~(\ref{eq:classical}), we need a slightly less general expression:
\begin{eqnarray}
\label{eq:specialresult} &&
\left\langle 
\Big[ D_z^{\ell} V(x,y,z) \Big]
\Big[ D_z^{\ell^{\prime}} V(x,y,z) \Big]
\right\rangle_C  
\nonumber \\ && =  (-1)^{\ell+\ell^{\prime}} 
\int_{0}^{\infty}\! dk \; W(k) k^{1+\ell+\ell^{\prime}} 
\exp [-2 k z] .
\nonumber\\
\end{eqnarray}
It is helpful to rewrite this in a dimensionless form.  
A natural length scale for the surface is $w=1/\sqrt{d}$, where
$d$ is the mean areal density of the surface patches.  
We assume that the covariance of the surface potential depends
on $w$ in such a way that it can be written in terms of a
scaled covariance function $\tilde{C}$ as
$C(\Delta r^\prime)=\Phi_{\rm rms}^2 \tilde{C}(\Delta r^\prime/w)$,  
where $\Phi_{\rm rms} \simeq [\langle V_s(x,y)^2 \rangle - \langle V_s(x,y) \rangle^2]^{1/2}$ is the rms variation of the surface potential from the mean.
We now introduce
\begin{eqnarray} 
\label{eq:scaledw}
\tilde{W}(u) \equiv && \int_{0}^{\infty} d(\Delta r^\prime/w) 
(\Delta r^\prime/w)  J_0(u \Delta r^\prime/w)
\tilde{C}(\Delta r^\prime/w), \nonumber \\ 
\end{eqnarray}
which allows us to rewrite Eq.~(\ref{eq:specialresult}) as
\begin{eqnarray} 
\label{eq:scaledbeforea}
&& \frac{\left\langle 
\Big[ D_z^{\ell} V(x,y,z) \Big]
\Big[ D_z^{\ell^{\prime}} V(x,y,z) \Big]
\right\rangle_C}
{(\Phi_{\rm rms}^2/w^{\ell+\ell^{\prime}})} \nonumber \\ && =
(-1)^{\ell+\ell^\prime} \int_{0}^{\infty}\! du \; \tilde{W}(u) \times u^{1+\ell+\ell^{\prime}} 
\times \exp [-2 u \: (z/w)] .
\nonumber\\
\end{eqnarray}
In general, for the $\tilde{W}(u)$ that we are interested in (see below), these integrals do not have closed forms.  
However, they may be approximated for large $z/w$ using an asymptotic technique. Part of the integrand, $\tilde{W}(u) \times u^{1+\ell+\ell^{\prime}}$,
may be written as a Taylor series in $u$ about $u=0$. Once multiplied with the rest of the integrand ($\exp [-2 u \: (z/w)]$),
the terms in the resulting series can be individually integrated in closed form (see, for example, Ref.~\cite{simmonds:1997}).  
Introducing $G(\ell+\ell^{\prime},z/w)$ as a shorthand for the left-hand side
of Eq.~(\ref{eq:scaledbeforea}), we obtain: 
\begin{eqnarray}
&&
G(L,z/w)= \nonumber\\ &&
(-1)^L \sum_{i=0,2,4...}
\frac{(L+1+i)!}{i!} \times \tilde{W}^{(i)}(0) \times 
\left(\frac{w}{2z}\right)^{L+2+i}, \nonumber\\
\end{eqnarray}
where $L=\ell+\ell^{\prime}$ and
$\tilde{W}^{(i)}(0)$ is the $i$th derivative of $\tilde{W}(u)$ evaluated
at $u=0$. Note that from its definition [Eq.~(\ref{eq:scaledw})],
the odd derivatives of $\tilde{W}(u)$ vanish at $u=0$.
For use later in this paper, we write out the first few terms of this series
for small $L$: 
\begin{subequations}
\label{eq:expansions}
\begin{eqnarray}
G(0,z/w)= &&
\frac{1}{4}   \tilde{W}(0)         
\left(\frac{w}{z}\right)^2 
+\frac{3}{16} \tilde{W}^{(2)}(0)   \left(\frac{w}{z}\right)^4
+ ... \nonumber\\
\\
G(1,z/w)= &&
-\frac{1}{4} \tilde{W}(0)       \left(\frac{w}{z}\right)^3 
-\frac{3}{8} \tilde{W}^{(2)}(0)          \left(\frac{w}{z}\right)^5
+ ... \nonumber\\
\\
\label{eq:g2asym}
G(2,z/w)= &&
\frac{3}{8} \tilde{W}(0)                 \left(\frac{w}{z}\right)^4 
+\frac{15}{16} \tilde{W}^{(2)}(0)         \left(\frac{w}{z}\right)^6
+ ... \nonumber\\
\\
G(3,z/w)= &&
-\frac{3}{4} \tilde{W}(0)                \left(\frac{w}{z}\right)^5 
-\frac{45}{16} \tilde{W}^{(2)}(0)        \left(\frac{w}{z}\right)^7
+ ... \; .\nonumber\\
\end{eqnarray}
\end{subequations} 

The first terms of these series are almost certain to dominate when $z\gg w$. 
From the definition of $\tilde{W}(u)$ in Eq. \ref{eq:scaledw}, it can be seen that $\tilde{W}(0)=0$ requires that the covariance function satisfies $C(\Delta r^\prime) < 0$ over some range of $\Delta r^\prime$, so that the integral taken over $\Delta r'$ is zero. This can be interpreted physically as antiferroelectric ordering of the surface potential; a case which seems unlikely to apply to polycrystalline metal surfaces.

It is important to note that, subject to the assumptions above, the details of $C(\Delta r^\prime)$ do not affect the $(z/w)$ scaling of $G(L,z/w)$ but only its magnitude. Therefore, the $z$ scaling of the patch fields is independent of the form of $C(\Delta r^\prime)$.

\section{Models for the surface patch potentials}
We will now calculate $G(L,z/w)$ using several different models for the electrostatic potential distribution on the surface. We start by calculating $C(\Delta r^\prime)$ for the model and then use this to find $\tilde{W}(0)$ and thus  $G(L,z/w)$.

A commonly used model for the surface potential covariance is of the
form \cite{rzchowski:1988,man:2006,dubessy:2009}
\begin{equation}
\label{eq:cmodel1}
C(\Delta r^\prime)=\Phi_{\rm rms}^2 
e^{- \gamma \left( \frac{\Delta r^\prime}{w} \right) },
\end{equation}
where $\gamma$ is dimensionless and on the order of 1.
This model follows from Poisson waiting statistics for grain boundary
crossings.  This, however, is an assumption, and a formal justification
does not appear in the literature.
An advantage of this model is that $\tilde{W}(u)$ has a closed form
[using Eq.~(\ref{eq:scaledw})]:
\begin{equation}
\tilde{W}(u)=\frac{\gamma}{[{\gamma}^2+u^2]^{3/2}},
\end{equation}
and thus the coefficients in the expansion of Eq.~(\ref{eq:expansions})
are readily determined 
[$\tilde{W}(0)=1/{\gamma}^2$, $\tilde{W}^{(2)}(0)=-3/{\gamma}^4$, etc.].

\begin{figure}
\includegraphics{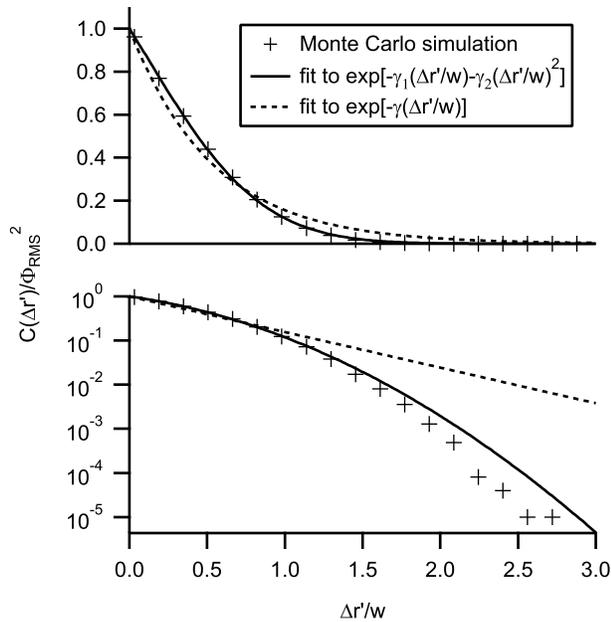}
\caption{\label{fg:montecarlo} Monte Carlo simulation of the surface
potential covariance function and two least-squares-fit models.
In the lower plot a logarithmic vertical axis is used to illustrate
the differences at large $\Delta r^{\prime}/w$.
}
\end{figure}

Motivated to provide a justification for 
Eq.~(\ref{eq:cmodel1}) (and determine a specific value for $\gamma$),
we performed Monte Carlo simulations 
to calculate a surface potential covariance function according
to the following recipe:
1) A total of $N$ patch ``centers'' were randomly put
within a square with sides of length $w\sqrt{N}$ (for a mean areal
patch density of $1/w^2$).  
2) At the center of this square, the patch with the closest center was determined.
3)  As we move out from the center of the square in a specific direction, eventually another
patch center becomes closer in distance than the initial one.  The point
at which this happens is considered to be
at a grain boundary, and beyond this point there is zero correlation 
between the local potential and the potential 
at the starting point in the center of the square.  
4) By
repeating this process (generating $N$ new patch centers within the
square, and traveling out from the center until a grain boundary is 
reached), we may
accumulate a surface potential correlation function.  
Provided $N$ is sufficiently
large, this model seems physically reasonable---we are assuming
that grains have grown isotropically outwards from randomly 
placed centers on a surface. Figure \ref{fg:montecarlo} illustrates the results of one of these 
Monte Carlo simulations. A least-squares fit to Eq.~(\ref{eq:cmodel1}) gives $\gamma \approx 1.9$, so that $\tilde{W}(0)\approx 0.28$.

We find that, instead of Eq.~(\ref{eq:cmodel1}), the covariance is a better fit to the relationship:
\begin{equation}
\label{eq:cmodel2}
C(\Delta r^\prime)=\Phi_{\rm rms}^2 
e^{
- \gamma_1 \left( \frac{\Delta r^\prime}{w} \right)
- \gamma_2 \left( \frac{\Delta r^\prime}{w} \right)^2
},
\end{equation}
with $\gamma_1 \approx 1.144(4)$ 
and $\gamma_2 \approx 0.993(6)$.  
The covariance falls off faster with increasing separation in this model.  Man et al.~\cite{man:2006} compared experimentally measured covariance
functions with a model similar to Eq.~(\ref{eq:cmodel1}) and also found that, although exponential decay was exhibited for small separations, the covariance falls off faster for increasing separations (see their Fig.~2).  However, a detailed comparison with our model is not possible as their surface was not isotropic.

\begin{figure}
\includegraphics{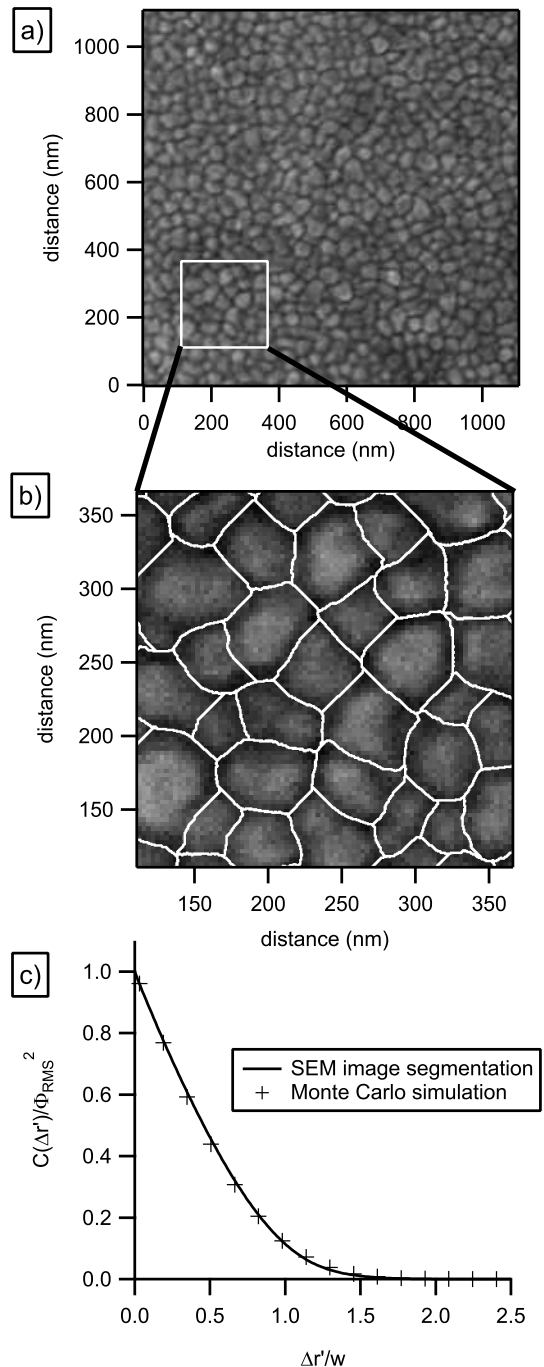}
\caption{\label{fg:semgrains}
(a) SEM image of a gold surface obtained by thermal evaporation (this is a portion of Fig.~4 of Ref.~\cite{cherry:2009}).
(b) Grain boundaries (indicated by white lines) over a small region of the image as determined by watershed segmentation.  The average area of a patch is $w^2=(44\unit{nm})^2$.
(c) Computed covariance of the surface potential based on segmentation of the SEM image.  To calculate this from the segmented image we assume constant surface potentials within  grains, and completely uncorrelated potentials between grains. The Monte Carlo simulation of Fig.~\ref{fg:montecarlo} is also shown for comparison.  
}
\end{figure}

We have tested this model by analyzing a scanning electron microscope (SEM) image of an evaporated gold structure on a silicon substrate (see Fig.~\ref{fg:semgrains}).   The ``watershed'' segmentation algorithm \cite{gonzalez:2002,sage:2010} was used to determine the location of the grain boundaries. To calculate the covariance function we assume that the potential measured at two points separated by $\Delta r^\prime$ is perfectly correlated if both points are on the same grain and uncorrelated if the points are on different grains.  As Fig.~\ref{fg:semgrains}(c) shows, the computed covariance is in good agreement with the Monte Carlo simulation [and thus also with the fit of Eq.~(\ref{eq:cmodel2})].

Unfortunately, a closed form for $\tilde{W}(k)$ does not appear to
be possible for the model of 
Eq.~(\ref{eq:cmodel2}).  Nonetheless, it is possible
to numerically compute the $\tilde{W}^{(i)}(0)$ required in Eq.~(\ref{eq:expansions}) for any $\tilde{C}$ using [see Eq.~(\ref{eq:scaledw})]:
\begin{equation}
\label{eq:numericalw}
\tilde{W}^{(i)}(0)=[D^{(i)}J_0(0)] \int_0^{\infty} d\alpha \:\:
\tilde{C}(\alpha)
\alpha^{1+i}.
\end{equation}
Performing these integrations for the model of Eq.~(\ref{eq:cmodel2}), we
find: $\tilde{W}(0)=\tilde{W}^{(0)}(0) \approx 0.207$,
and $\tilde{W}^{(2)}(0) \approx -0.064$.

\section{Fluctuations in the electric field}

The $G(2,z/w)$ function determines the variance of the
patch electric fields:  
$\langle F_z^2 \rangle_C \approx (3/8) \times (\Phi_{\rm rms}^2 / w^2) \times
\tilde{W}(0) \times (w/z)^4$.  Using Eq.~(\ref{eq:mainresult}),
the variances of 
the $x$ and $y$ components of the
electric field can be calculated.  We find that they are
each 1/2 of the result for $z$.  
Thus, we can summarize;
with the model of Eq.~(\ref{eq:cmodel2}),
the rms electric field for $z \gg w$ is
\begin{equation}
\label{eq:thebigresult}
E_{\rm rms} \approx 0.39 \frac{\Phi_{\rm rms}}{w} \left( \frac{w}{z} \right)^2.
\end{equation}
This result is not especially sensitive to the particular patch model.  
For example, we have performed numerical simulations of the patch field over a large array of square patches (each $w$ by $w$) with random potentials  at distances $z\gg w$ and found that the numerical prefactor in Eq.~(\ref{eq:thebigresult}) is $0.33$ instead of $0.39$. The model of Eq.~(\ref{eq:cmodel1}) gives a numerical prefactor of 0.46.
An approximate estimate similar to Eq.~(\ref{eq:thebigresult}) has been provided by Sandoghdar {\it et al.}~\cite{sandoghdar:1996} and used by Mozley {\it et al.}~\cite{mozley:2005}.

\begin{figure}
\includegraphics{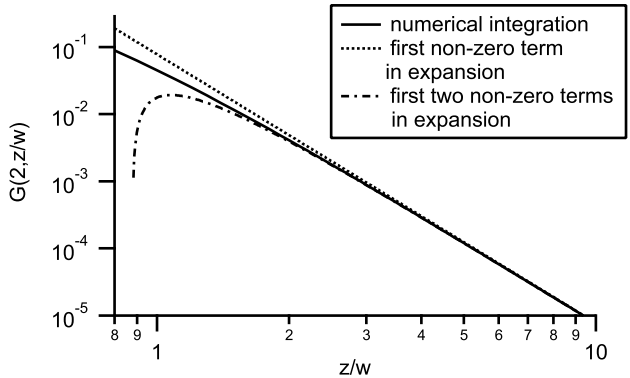}
\caption{\label{fg:validex} 
Comparison between numerical calculation of $G(2,z/w)$ 
for the model of Eq.~(\ref{eq:cmodel2})  
(using Eq.~(\ref{eq:specialresult})), and the asymptotic series expansion
of Eq.~(\ref{eq:g2asym}), with the $\tilde{W}^{(i)}(0)$ 
coefficients calculated using Eq.~(\ref{eq:numericalw}).
}
\end{figure}

It is worth asking when the higher-order terms of Eq.~(\ref{eq:expansions})
can be neglected.  In Fig.~\ref{fg:validex}, we calculate $G(2,z/w)$ by direct integration of Eq.~(\ref{eq:specialresult}).  The results due to the first two terms of Eq.~(\ref{eq:g2asym}) are also shown. The figure indicates that  keeping only the first term is an excellent approximation for $z \gg w$ (a similar plot for the model of Eq.~(\ref{eq:cmodel1}) is given in Fig.~1 of Dubessy {\it et al.}~\cite{dubessy:2009}).

The rms patch field and the image field of an elementary-charge both scale in the same way with distance to the surface, so it is interesting to compare their magnitudes.  If we assume a potential fluctuation of $\Phi_{\rm rms}^2=(90 \unit{mV})^2$ and $w = 50 \unit{nm}$, typical of thermally evaporated gold surfaces \cite{PhysRevLett.90.160403,groth:2980}, we find that the rms electric field due to patches is approximately 5 times that of the elementary charge image field: $|\vec{E}_i|=q_e/[4\pi\epsilon_0 (2z)^2]$.    Seeing the influence of the image field due to an elementary charge near such a surface would be difficult.

Despite its simplicity and intrinsic importance, there does not appear to 
be any clear experimental observations that would support 
the validity of Eq.~(\ref{eq:thebigresult}). Initial experiments with Rydberg atoms using microtrap technology have observed large dc fields due to the deposition of Rb on the surface \cite{PhysRevA.81.063411} (see also Ref.~\cite{PhysRevA.75.062903}), possibly masking the influence of patch fields.  Alkali adsorption has been recognized as a problem since the very early days of Rydberg atom-surface interaction experiments \cite{Kocher198768}.  Some theoretical work on the influence of adsorbates has been done in the context of ion-surface collisions \cite{Miskovic1995431}. To avoid the problem of adsorbates, Dunning's group switched to using xenon Rydberg atoms in their surface studies \cite{PhysRevLett.85.5444}.  

Dunning's group has recently studied Rydberg atom image field ionization using Au(111) samples \cite{pu:2010}. The surfaces consisted of multiple grains, typically 300--500 nm in size. Possibly due to contamination,
the surface potential was inhomogeneous, with variations of up to 70 mV from the average over length scales of 50--250 nm (shown in Fig.~1 of Ref.~\cite{pu:2010}).  By using a scanning Kelvin probe to measure surface potential, they computed the statistical properties of the electric field above the surface and found that the field is consistent with their observed image field ionization.  For $z \gtrsim 60$ nm, they found $E_{\rm rms} \approx (5 \times 10^{-10} \unit {V \; m})/z^2$.  
Assuming our polycrystalline model is applicable, with $\Phi_{\rm rms} \approx 22\unit{mV}$ (determined from our analysis of Fig.~1 of Ref.~\cite{pu:2010}) and $w=100\unit{nm}$, Eq.~(\ref{eq:thebigresult}) predicts slightly larger fields, with $E_{\rm rms} \approx (9 \times 10^{-10} \unit {V \; m})/z^2$. Given the uncertainty in determining $w$ from the figure, this is reasonably consistent with the result in Ref.~\cite{pu:2010}.

It is possible that polar or polarizable contaminants at grain boundaries could {\em reduce} the magnitude of the patch fields.  Darling \cite{darling:1989} did extensive scanning Kelvin probe measurements of the work function immediately above copper surfaces with large grain sizes, found that fluctuations were significantly less than one might expect, and attributed the reduction to oxidation of the surface and physisorbed molecules (e.g.~water).

\section{Patch fields and Rydberg atoms -- estimates of energy level shifts}

The statistical properties of the patch fields may now be combined with the atomic properties to predict the variance in the energy levels using Eq.~(\ref{eq:classical}).  Writing this as a series in $w/z$:
\begin{eqnarray} 
\label{eq:fieldandatom}
\langle (\Delta E)^2 \rangle_{C} \; 
 \approx && \: (M_1)^2 \langle F_z^2 \rangle_C + 2 M_1 M_2 
\langle F_z \partial_z F_z \rangle_C  + ... \nonumber\\ 
 \approx && \: (M_1)^2 \times (3/8)\tilde{W}(0) (\Phi_{\rm rms}^2/w^2) 
\left(\frac{w}{z}\right)^4 \nonumber \\
&& \: + 2 M_1 M_2 \times (-3/4)\tilde{W}(0) 
(\Phi_{\rm rms}^2/w^3) \left( \frac{w}{z} \right)^5 
\nonumber\\
&& \: + O(\left(\frac{w}{z}\right)^6). \nonumber\\
\end{eqnarray}

The first term in this expansion is due to the rms $z$ field and the atom's electric dipole, and is expected to be dominant at large $z$.  However,
the higher order terms in Eq.~(\ref{eq:fieldandatom}) can be enhanced relative
to the lower-order terms by increasing $n$ (increasing the size of the atom).  
As the classical outer turning point of the Rydberg electron is $\approx n^2$ (in atomic units), the multipole moments of order $\ell$ scale with
$n$ like $M_{\ell} \approx n^{2\ell}$ (see, for example, Ref.~\cite{gallagher:1994}). These higher-order multipoles sense the
field {\em variations} over the Rydberg orbit.

Under what conditions will variations in the patch fields over the extent of {\em individual} atoms contribute to the inhomogeneous broadening?  We may estimate this by equating the first two terms written explicitly in Eq.~(\ref{eq:fieldandatom}). This tells us that the size of the Rydberg atom, $n^2$, has to be approximately the distance of the atom to the surface before these would be comparable.  Due to the interaction of the Rydberg atom with its image, this is a highly nonperturbative situation \cite{chaplik:1968}. 
We conclude that it would be difficult to observe any effect of the variation in patch fields over the orbits of individual Rydberg atoms (at least when they have dipole moments of order $n^2$).  An additional qualitative justification is given in the Appendix.

We now give a simple numerical estimate for the inhomogeneous broadening of Rydberg energy levels due to patch fields.  When $z$ is large compared to $w$ and the atom size, the first term in Eq.~(\ref{eq:fieldandatom}) dominates and the rms broadening will be $\delta E = M_{1} \sqrt{\langle F_z^2 \rangle_{C}}$.
For the extreme Stark states of hydrogen we have $M_{1} = \mu_z = (3/2) n (n-1)$, which for $n=30$ is $1.7 \unit{GHz/(V/cm)}$  (this is also reasonable for non-hydrogenic atoms, assuming a large
enough dc field is applied).  For a typical thermally evaporated gold surface \cite{PhysRevLett.90.160403,groth:2980}, 
we assume $\Phi_{\rm rms}^2=(90 \unit{mV})^2$ and $w = 50 \unit{nm}$,
giving $\sqrt{\langle F_z^2 \rangle_{C}} \approx 0.13 \unit{V/cm}$.   We find that $\delta E \approx 200 \unit{MHz}$ -- which should
be straightforward to observe in optical excitation.  
For Rb atoms, a possible spectroscopic probe would be the last step in
the $5s \rightarrow 5p \rightarrow 5d_{5/2} \rightarrow nk$ 
excitation sequence (where the last transition is enabled by a dc field sufficient to mix $f$ character into the reddest $nk$ states). 

The extreme Stark states provide the largest broadening.  Broadening due to patch fields will be much lower than this estimate if low angular momentum states are excited at fields small enough so that the Stark effect is second order.

\section{Summary and Outlook}

Rydberg atoms with permanent electric dipole moments have a high sensitivity to electric fields.  We have shown that the patch fields near a typical metal surface can be large compared to the image field of an elementary charge and should be expected to cause measurable inhomogeneous broadening of Rydberg energy levels.  The rms spatial variation in the field strength has a distance dependence of $1/z^2$. Spatial variations in the fields over the Rydberg atom orbit do not appear to be important.  An experiment to verify the magnitude of the rms field and the expected scaling with surface distance [see Eq.~(\ref{eq:thebigresult})] would be useful in assessing the feasibility of coherently manipulating Rydberg atoms near polycrystalline surfaces and in planning future experiments. 

\acknowledgments
We gratefully acknowledge discussions with Chris Gray (Guelph), and thank T. Darling (University of Nevada, Reno) for providing a copy of his PhD thesis \cite{darling:1989}, 
which contains useful background information on patch effects.  We thank M. Mazurek and C. Liekhus-Schmaltz for comments on this manuscript.  This work was supported by NSERC.  

\appendix

\section{Potential and field covariance functions}

There is a qualitative way to understand why the inhomogeneities in the patch fields over the extent of individual Rydberg atoms would be difficult to observe.  Equation (\ref{eq:mainresult}) can be used to determine the covariance between the potential and derivatives of the potential measured at different locations in space.  Using the result
for the 2D Fourier transform of a radially symmetric function 
we obtain:
\begin{eqnarray}
\label{eq:correl}
&& \left\langle 
\Big[ D_z^{\ell} V(x,y,z) \Big]
\Big[ D_z^{\ell^{\prime}} V(x+\Delta x, y + \Delta y,  z+\Delta z) \Big]
\right\rangle_C  \nonumber\\&& = 
(-1)^{\ell+\ell^{\prime}} \frac{\Phi_{\rm rms}^2}{w^{\ell+\ell^{\prime}}} 
\times \int_{0}^{\infty}\! du \; 
\tilde{W}(u) u^{1+\ell+\ell^{\prime}}  \nonumber \\ && 
\times J_0(u\Delta r/w) \exp [-u (2z+\Delta z)/w], \nonumber \\
\end{eqnarray}
where $\Delta r = \sqrt{ \Delta x^2 + \Delta y ^2}$.

Again, like with Eq.~(\ref{eq:scaledbeforea}),
this integral can be approximated for large $z/w$ by writing the
$\tilde{W}(u) u^{1+\ell+\ell^{\prime}}$ part of the integrand
as a Taylor series and then integrating the individual terms.
For covariances in the potential, we obtain
for the first nonzero term
\begin{eqnarray}
&& \left\langle 
 V(x,y,z) V(x+\Delta x, y + \Delta y, z+\Delta z) 
\right \rangle_C  \nonumber\\&& \approx
\frac{1}{4}  \tilde{W}(0) \Phi_{\rm rms}^2 
\left( \frac{w}{z} \right) ^2
\frac{1}{\left(1+\frac{\Delta z}{2z}\right)^2}
\frac{1}{\left[1+\left( \frac{\Delta r}{2z+\Delta z} \right)^2 \right]^{3/2}}.
\nonumber\\
\end{eqnarray}
For covariance in the z-component of the electric field, we obtain for the
first nonzero term
\begin{eqnarray}
&& \left\langle 
\Big[ D_z^1 V(x,y,z) \Big]
\Big[ D_z^1 V(x+\Delta x, y + \Delta y, z+\Delta z) \Big]
\right\rangle_C  \nonumber\\&& \approx
\frac{3}{8} 
\frac{\tilde{W}(0) \Phi_{\rm rms}^2}{w^2} 
\left( \frac{w}{z} \right) ^4
\frac{1}{\left(1+\frac{\Delta z}{2z}\right)^4}
\frac{\left[ 1-\frac{3}{2}\left(\frac{\Delta r}{2z+\Delta z} \right)^2 \right] }
{\left[1+\left( \frac{\Delta r}{2z+\Delta z} \right)^2 \right]^{7/2}}.
\nonumber\\
\end{eqnarray}
Higher order terms involve larger powers of $w/z$.  These results have
been written in a way to emphasize the influence of nonzero 
$\Delta r$ and $\Delta z$ as a correction factor to the 
$\Delta r = \Delta z = 0$ result.  It is apparent that $z$, the 
distance to the surface, sets the length scale for spatial variations
in the potential and fields.  Thus, we can understand in a qualitative way 
the results of the main text:  a Rydberg atom should have a size
comparable to its distance from the surface for spatial variations to 
be significant.

If an atom is moving near a surface, spatial variations in the fields manifest themselves as time-dependent variations experienced in the atom's frame.  In this case, we note that the calculations of this section could be adapted to determine the power spectral densities of these fluctuations (using
the Wiener-Khinchin theorem). 

\bibliography{references}

\end{document}